\documentstyle[aps,epsf]{revtex}

\title{An Excitation Function of K$^-$ and K$^+$
Production in Au+Au Reactions at the AGS
\footnote{Corresponding Author: C.A. Ogilvie, 
Department of Physics and Astronomy, Iowa State University, 
Ames, IA 50011-3160,
cogilvie@iastate.edu}}

\begin{document}
\input epsf
\maketitle

\renewcommand{\thefootnote}{\alph{footnote}}
\begin{center}
\it{E866 Collaboration}
\end{center}

L. Ahle$^{10}$\footnote{Lawrence Livermore National Laboratory, Livermore CA 94550}, 
Y.~Akiba$^{6}$, K.~Ashktorab$^{2}$, 
M.D.~Baker$^{10}$\footnote{Brookhaven National Laboratory, Upton, NY 11973}, D. Beavis$^{2}$,
B.~Budick$^{11}$, J.~Chang$^{3}$, C.~Chasman$^{2}$, Z.~Chen$^{2}$, 
Y.Y.~Chu$^{2}$, T.~Chujo$^{15}$, 
J.~Cumming$^{2}$, R.~Debbe$^{2}$, J.C.~Dunlop$^{10}$\footnote{Yale University, New Haven, CT 06520},  
W.~Eldredge$^{3}$,
K.~Fleming$^{3}$,
S.-Y.~Fung$^{3}$, E.~Garcia$^{9}$, S.~Gushue$^{2}$,  H.~Hamagaki$^{14}$,
R.~Hayano$^{13}$,  G.H.~Heintzelman$^{10b}$, 
J.H.~Kang$^{16}$, E.J.~Kim$^{2,16}$,  A.~Kumagai$^{15}$,
K.~Kurita$^{15}$, J.H.~Lee$^{2}$, Y.K.~Lee$^{8}$, Y.~Miake$^{15}$, 
A.C.~Mignerey$^{9}$, 
M.~Moulson$^{4}$\footnote{Laboratori Nazionali di Frascati, INFN, 00044 Frascati, Italy}, 
C.~Muentz$^{2}$\footnote{Goethe Universit\"at, Institut f\"ur Kernphysik, Frankfurt, Germany}, 
K.~Nagano$^{13}$, C.A.~Ogilvie$^{10}$, J.~Olness$^{2}$,  K.~Oyama$^{13}$, 
L.~Remsberg$^{2}$,  H.~Sako$^{15}$, 
R.~Seto$^{3}$,  J.~Shea$^{9}$,  K.~Shigaki$^{7}$, 
S.G.~Steadman$^{10}$\footnote{U.S. Department of Energy, Office of High Energy and Nuclear Physics
SC-23, Germantown, MD 20874}, 
G.S.F.~Stephans$^{10}$,  T.~Tamagawa$^{13}$,
M.J.~Tannenbaum$^{2}$,  S.~Ueno-Hayashi$^{15}$,
F.~Videbaek$^{2}$, H.~Xiang$^{3}$, F.~Wang$^{4}$\footnote{Purdue University,
West Lafayette, IN 47907}, K.~Yagi$^{15}$, 
F.~Zhu$^{2}$,\\

\begin{center}
\it{E917 Collaboration}
\end{center}

B.B.~Back$^{1}$, R.R.~Betts$^{1,5}$, J.~Chang$^{3}$,
W.C.~Chang$^{3}$\footnote{Institute of Physics, Academica Sinica, Taipei 11529, Taiwan}, 
C.Y.~Chi$^{4}$, Y.Y.~Chu$^{2}$, J.B.~Cumming$^{2}$,
J.C.~Dunlop$^{10c}$, W.~Eldredge$^{3}$, S.Y.~Fung$^{3}$, 
R.~Ganz$^{5}$,
E.~Garcia$^{9}$, A.~Gillitzer$^{1}$\footnote{Institut f\"ur Kernphysik, 
Forschungzentrum J\"ulich,D-52425 J\"ulich, Germany}, G.H.~Heintzelman$^{10b}$,
W.F.~Henning$^{1}$\footnote{Gesellschaft f\"ur Schwerionforschung, D-64291 
Darmstadt, Germany}, D.J.~Hofman$^{1,5}$, B.~Holzman$^{5}$, J.H.~Kang$^{16}$,
E.J.~Kim$^{2,16}$, S.Y.~Kim$^{16}$, Y.~Kwon$^{16}$,
D.~McLeod$^{5}$, A.C.~Mignerey$^{9}$, M.~Moulson$^{4,d}$,
V.~Nanal$^{1}$\footnote{Tata Institute of Fundamental Research, Colaba, Mumbai 400005 India}, 
C.A.~Ogilvie$^{10}$, R.~Pak$^{12}$,
A.~Ruangma$^{9}$, D.~Russ$^{9}$, R.~Seto$^{3}$, P.J.~Stanskas$^{9}$,
G.S.F.~Stephans$^{10}$, H.~Wang$^{3}$, F.L.H.~Wolfs$^{12}$, 
A.H.~Wuosmaa$^{1}$,
H.~Xiang$^{3}$, G.H.~Xu$^{3}$, H.B.~Yao$^{10}$, C.M.~Zou$^{3}$
\\
\begin{center}
{\it
$^{1}$ Argonne National Laboratory, Argonne, IL 60439
\\
$^{2}$ Brookhaven National Laboratory, Upton, NY 11973
\\
$^{3}$ University of California Riverside, Riverside, CA 92521
\\
$^{4}$ Columbia University, Nevis Laboratories, Irvington, NY 10533
\\
$^{5}$ University of Illinois at Chicago, Chicago, IL 60607
\\
$^{6}$ High Energy Accelerator Research Organization 
(KEK), Tanashi-branch, (Tanashi) Tokyo 188, Japan\\
$^{7}$ High Energy Accelerator Research Organization 
(KEK),Tsukuba, Ibaraki 305, Japan\\
$^{8}$ Johns Hopkins University, Baltimore, MD 
\\
$^{9}$ University of Maryland, College Park, MD 20742
\\
$^{10}$ Massachusetts Institute of Technology, Cambridge, MA 02139
\\
$^{11}$ New York University, New York, NY 
\\
$^{12}$ University of Rochester, Rochester, NY 14627
\\
$^{13}$ Department of Physics, University of Tokyo, Tokyo 113, Japan 
\\
$^{14}$ Center for Nuclear Study, School of Science, University of Tokyo, 
Tanashi, Tokyo 188, Japan\\
$^{15}$ University of Tsukuba, Tsukuba, Ibaraki 305, Japan \\
$^{16}$ Yonsei University, Seoul 120-749, South Korea
\\
}
\end{center}
\begin{abstract}

Mid-rapidity spectra and yields of K$^-$ and K$^+$ have been measured
for Au+Au collisions
at 4, 6, 8, and 10.7~AGeV.
The K$^-$ yield increases faster 
with beam energy  than for K$^+$ and hence the K$^-$/K$^+$ ratio increases
with beam energy.
This ratio is studied as a function of both $\sqrt{s}$ and
$\sqrt{s}$-$\sqrt{s_{th}}$
which allows the direct comparison of the 
kaon yields with respect to the production threshold in p+p reactions.
For equal $\sqrt{s}$ - $\sqrt{s_{th}}$ the measured ratio
K$^-$/K$^+$=0.2  at energies above threshold in contrast to the 
K$^-$/K$^+$ ratio of near unity observed at energies below threshold. The use
of the K$^-$/K$^+$ ratio to 
test the predicted changes of kaon properties 
in dense nuclear matter is discussed.

\end{abstract}


In many-body physics  the effects of  multiple
two-body interactions are often incorporated
into models as a mean-field which governs the propagation of single particles.
In nuclear matter the mean-field for nucleons and $\Lambda$ hyperons is
well-established, but less is known about the mean-field 
experienced by mesons\cite{Dover,Yam98}, especially strange mesons. 
From K$^+$+p scattering experiments,  
the two-body interaction between K$^+$ and baryons is known to be
repulsive\cite{Dover,kpscatter}, leading to the expectation of a repulsive
mean-field for K$^+$\cite{Dover,Weise96,Li96,Ehe96,Schaf97}. 
In contrast, studies of K$^-$-atoms show that the mean-field interaction of K$^-$
with baryons is attractive\cite{kmatoms}.
The opposite-sign interaction for K$^+$ and K$^-$ may make it
possible to use the differences between K$^+$ and K$^-$ 
production  as a
probe of the mean-fields experienced by mesons during 
nuclear reactions.

An alternative method of treating 
the many two-body interactions a kaon experiences is 
via the dispersion relation between its momentum and 
energy in nuclear matter\cite{Weise96,Schaf97}. 
At zero-momentum the energy is interpreted as an 
effective mass, such that
the repulsive K$^+$-baryon interactions increase the
effective mass of K$^+$ and the attractive K$^-$-baryon interactions
decrease the effective mass of 
K$^-$\cite{Weise96}. The predicted  changes in mass are large,
with a 50\% decrease in  K$^-$ effective mass at three times
nuclear matter density\cite{Li96,Ehe96,Schaf97}.
In most calculations\cite{Weise96,Schaf97} 
the decrease in the K$^-$ effective mass is
larger than the increase for the K$^+$ mass. Hence the 
available phase space and
therefore the yields of 
K$^+$K$^-$ pair production could increase inside a dense medium.
The theoretical issues are far from certain,
e.g., for kaons at finite momentum the K$^-$ potential is predicted 
to change from attractive to repulsive\cite{JSB99} .

Experimentally the yield of K$^-$ from Ni+Ni at 1.85 and 1.66 AGeV 
has been measured 
by the FRS collaboration (GSI)\cite{FRS94}, and 
K$^-$ and K$^+$ yields from  Ni+Ni collisions 
between 0.8 and 1.8~AGeV have been measured 
by the KaoS collaboration (GSI)\cite{kaosNiNi}.
Attempts have been made to describe these data with various
transport models\cite{Li94,Cass97,JSB99}.
If free kaon masses are included in these models,
then the calculated yields 
underpredict the K$^-$ data 
by a factor of five. 
However, if either in-medium 
kaon masses\cite{Li94,Cass97} or density-dependent 
kaon production cross-sections\cite{JSB99} are used in the calculations
then the measured K$^-$ yields can be reproduced.
 
Since the yield of particles indirectly probes 
the in-medium properties of hadrons, the
transport models must be confronted with the
widest possible range of K$^+$ and K$^-$ 
production data  as a critical check of consistency.
It is difficult to distinguish in-medium effects from other
hadronic processes that affect kaon production.
For example, secondary collisions often occur between hadrons in 
resonant states, and the excitation energy of the resonances is
then available for particle production.
At a beam energy of 10.7~AGeV
the kaon yield per collision participant
in central Au+Au reactions\cite{kpkm}
has been measured to be three to four times that for
p+p reactions at the same energy, consistent with the majority of
kaon production coming from secondary rather than primary
nucleon-nucleon collisions.
At lower beam energies, a larger fraction 
of these secondary collisions will be below
the threshold for producing a pair of strange hadrons.
In models of nucleus-nucleus reactions at 1-2~AGeV\cite{Li94,Cass97},
kaon production is dominated by
the rare  secondary collisions that
are above threshold compared to production from 
primary collisions boosted by Fermi motion.
Given their likely dominant role in kaon production, the detailed
effects of secondary collisions need to be fully
understood before possible in-medium effects can be firmly established.

In this Letter we report on the measurement of K$^-$ production at 
4, 6, 8 and 10.7~AGeV Au+Au collisions and
compare it with published K$^+$ data at the same 
beam energies and also at 2 AGeV\cite{kpkm,Back99}. 
The data come from two separate AGS experiments at BNL, E866 and 
E917; each of which used much the
same apparatus.  The E866 collaboration measured
Au+Au collisions at 1.96, 4.00, and 10.74~AGeV 
kinetic energy. 
Experiment E917 measured Au+Au reactions at 5.93 and 7.94~AGeV 
kinetic energy. The beam energies quoted and used in this Letter correspond 
to the energy half-way through the target.
For shorthand these will be referred to 
in the text as 2, 4, 6, 8 and 10.7 AGeV.

The new data presented in this
Letter (K$^-$ from Au+Au at 4, 6, and 8~AGeV)
were measured with the Henry Higgins 
spectrometer used by experiments E802, E859, E866 and E917 of the BNL AGS.
For more details on the equipment and data analysis the
reader is referred to references\cite{kpkm,Chen98,nim} and only
a brief description is
provided here.
An Au target of 1960 mg/cm$^2$ was used for each measurement which 
corresponds to approximately a 3\% interaction probability for the Au projectile.
The Au beam loses approximately 0.06 AGeV as it passes through the target.
The rotatable  spectrometer has an
acceptance of 25~msr with an opening angle 
$\Delta\theta\sim 14^o$. It consists of drift and multi-wire chambers placed 
on either side of a dipole magnet. 
Similar track reconstruction, efficiency corrections, acceptance and decay
algorithms were used for the data at each
beam energy.
Particle identification was performed using
time-of-flight scintillators with a timing resolution of 130~ps.  
The detectors for global event characterization were
a multiplicity array surrounding the target and a calorimeter 
with a 1.5$^o$ opening angle placed 
at zero degrees.
For the 4, 6, and 8~AGeV Au+Au data analysis, 
central collisions were selected with the multiplicity
array.
The data at 10.7~AGeV  were measured prior to installation of 
the multiplicity detector; so, 
central collisions at this beam energy were selected using
the zero-degree calorimeter.
A comparison of event selection using these two devices has been 
published\cite{double}. 

Based on similar analyses fully described in 
references\cite{kpkm,Chen98}
the total systematic uncertainty in the  
normalization of the particle yields is 15\% and is 
dominated by the uncertainty in the acceptance, the tracking efficiency, 
and the extrapolation to low transverse 
mass required to obtain total particle yields.
The systematic uncertainty on the mean transverse mass
extracted from the spectra is estimated to be 10\%.
The relative beam-energy to beam-energy
systematic uncertainty in both the yields and mean transverse mass
is 5\% because the tracking efficiency depends on 
the occupancy of the
spectrometer\cite{kpkm}, and the uncertainties 
involved in extrapolating to obtain particle yields are not the same for each
beam energy.  

\begin{figure}[htb]
\epsfxsize=18cm\epsfbox[0 0 500 300]{
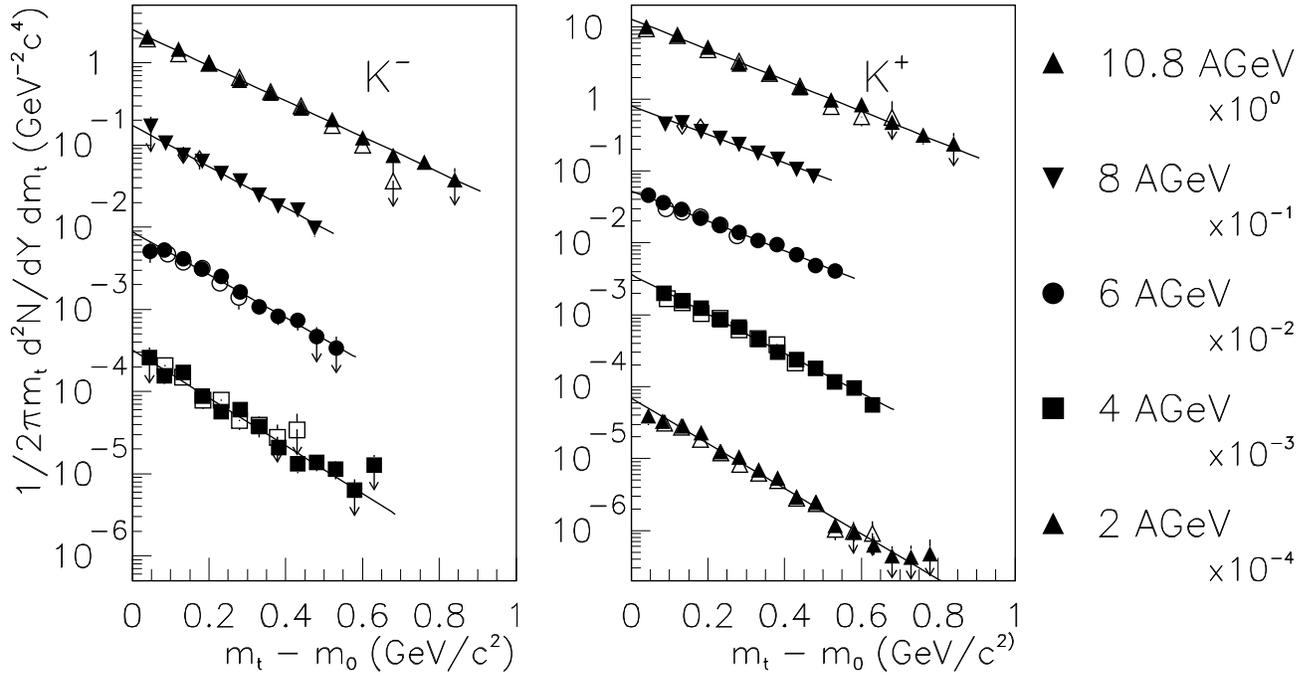}

\caption{
The invariant yields of kaons plotted as a function of m$_t$
from the most central 5\% Au+Au reactions at  2 (K$^+$ only), 
4, 6, 8, 10.7 AGeV. 
The left panel shows K$^-$ spectra the  right panel
shows K$^+$.
For each energy the spectrum just below mid-rapidity is shown with 
filled symbols,
the spectrum just above mid-rapidity is shown with open symbols.   
The data at 2, 4, and 10.7 AGeV are from E866 and the 
data at 6, and 8 AGeV are from
E917. The arrows indicate that the error bar continues to zero 
cross-section. The data at 10.7~AGeV are shown on the
correct scale, the data at each lower energy are 
divided by successive powers of ten for clarity.
The plotted errors are only statistical.}
\label{fig:mtkaon0p5}
\end{figure}

The invariant yields of kaons produced in central Au+Au collisions
(corresponding to the most central 5\% of the total interaction 
cross-section; $\sigma_{int}$=6.8b) are plotted in Figure \ref{fig:mtkaon0p5}
as a function of transverse mass, m$_t=\sqrt{p_t^2+m_0^2}$, 
for beam energies of 2, 4, 6, 8, and 10.7~AGeV, 
where $p_t$ is the transverse momentum of the kaon and $m_0$ its rest mass.
Due to low statistics there is no K$^-$ spectrum at 2 AGeV.
For beam energies of 2, 4, 6, and 8~AGeV there are
two m$_t$ spectra shown in Figure
\ref{fig:mtkaon0p5}. 
One corresponds to a rapidity slice just below mid-rapidity,  
$-0.25<\frac{y-y_{nn}}{y_{nn}}<0$ (shown as filled symbols in 
Figure \ref{fig:mtkaon0p5}) and  the other is
just above mid-rapidity,
$0<\frac{y-y_{nn}}{y_{nn}}<0.25$ 
(shown as open symbols), 
where $y_{nn}$ is mid-rapidity in the laboratory frame.
The kaon spectra at 10.7~AGeV have already been published\cite{kpkm}
and correspond to rapidity bin-widths of half this size. 
Combining two rapidity bins at 10.7~AGeV at mid-rapidity 
reduces the extracted dN/dy values by less than 5\%.
 
Each pair of m$_t$ spectra 
covers a symmetric rapidity region around mid-rapidity 
and the two spectra are statistically consistent with each other in all cases,
so a single exponential function in m$_{t}$ was 
fit simultaneously to the pairs spectra: 
\begin{equation}
\frac{1}{2\pi m_t}\frac{d^2N}{dm_tdy}=
\frac{dN/dy}{2\pi(Tm_0 + T^2)}e^{-(m_t-m_0)/T} \hspace{0.5in}. 
\end{equation}
The fits reproduce the spectra well with two free parameters,
the inverse slope
parameter T and the rapidity density, dN/dy, in that
rapidity slice. These parameters
are tabulated in Table~\ref{tabk0p5}
along with the mean m$_t$, $<m_t>$, calculated from the
inverse slope parameter. These values incorporate an extrapolation
to both low and high regions of m$_t$ where the experiment has no
acceptance. The extrapolations contribute
less than 20\% to the total yield, hence the systematic
uncertainty on dN/dy and $<m_t>$ arising 
from the extrapolation is estimated to be less than 5\%. 
 
The values of $<m_t>$ for both K$^+$ and K$^-$ 
increase with beam energy from 4 to 6 AGeV. Above 6 AGeV the
changes in $<m_t>$ are approximately comparable to  
the estimated beam-energy to beam-energy 
systematic uncertainty of 5\%. 
Throughout the energy range, 
the $<m_t>$ for K$^+$ is higher than the $<m_t>$ for 
K$^-$ by approximately 10-70 MeV/c$^2$.
Part of this difference could be attributed to the Coulomb
interaction\cite{kapusta}. Coulomb effects on spectra require modelling
the dynamical expansion of the source and the space-time distribution
of kaon production.
Any difference in the kaon spectra beyond Coulomb effects
would challenge the assumptions made in  global-equilibrium 
thermal models\cite{Chap,Schned,PBM,Cley}. These models
assume that the system is at kinetic equilibrium so that the spectra of equal 
mass particles (e.g. K$^+$ and K$^-$) should be identical in shape.
It has also been predicted\cite{Song99} that K$^-$ are attracted to lower
values of m$_t$ by an attractive mean-field thereby lowering the K$^-$
$<m_t>$.

\begin{figure}[htb]
\begin{minipage}[t]{175mm}
\epsfxsize=17.5cm\epsfbox[10 0 670 275]{
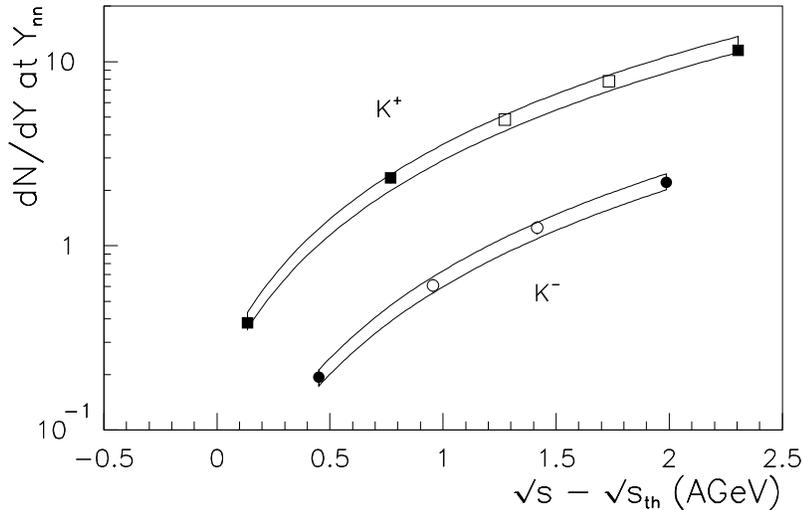}

\end{minipage}
\caption{
The mid-rapidity yields of K$^+$ and K$^-$ as a function of 
$\protect\sqrt{s}$-$\protect\sqrt{s_{th}}$  
for the most central 5\% Au+Au reactions.  
The data from E866 are shown as filled symbols and the data from
E917 are shown as open symbols.
The K$^+$ data have been published in ref [14,15].  
The error bars include both statistical uncertainty and the 5\% relative systematic uncertainty.
The fitted lines are used to interpolate between the data
points.}
\label{fig:yieldexcit}
\end{figure}

The mid-rapidity yields of $K^+$ and 
K$^-$ are shown as a function of 
$\sqrt{s}$-$\sqrt{s_{th}}$ (AGeV) 
in Figure~\ref{fig:yieldexcit}, where $\sqrt{s_{th}}$ is the minimum
energy
needed to produce a kaon in a p+p reaction ($\sqrt{s_{th}}$ =
2.548 AGeV for K$^+$ production and $\sqrt{s_{th}}$= 2.864~AGeV for
K$^-$ production).  
The measured kaon yields in Au+Au increase steadily with beam energy
and have been fit with a second order polynomial in 
$x = \sqrt{s} - \sqrt{s_{th}}$ in order to
interpolate between the measured data points.

\begin{figure}[htb]
\begin{minipage}[t]{175mm}
\epsfxsize=17.5cm\epsfbox[10 0 670 275]{
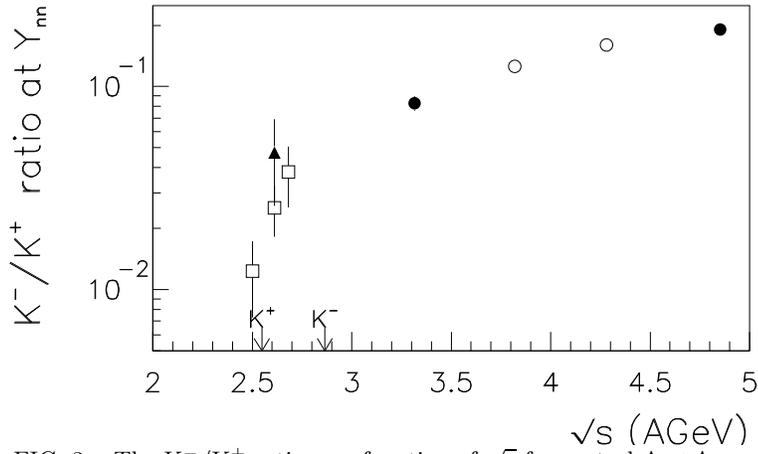}

\end{minipage}

\caption{
The K$^-$/K$^+$ ratio as a function of $\protect\sqrt{s}$
for central Au+Au reactions.  
The data from E866 are shown as filled circles and the data from
E917 are shown as open circles. 
Also shown  are the K$^-$/K$^+$ ratios measured by the KaoS collaboration
in  Ni+Ni (triangles)\protect\cite{kaosNiNi} 
and C+C (open squares)\protect\cite{kaosCC}.}
\label{fig:ratioexcit}
\end{figure}

The K$^-$/K$^+$ ratio for central
reactions is shown in 
Figure~\ref{fig:ratioexcit} plotted as a function of $\sqrt{s}$. Also shown in
this figure are the K$^-$/K$^+$ ratios measured by the KaoS collaboration in 
central Ni+Ni\cite{kaosNiNi} and minimum-bias
C+C\cite{kaosCC} reactions at lower beam energies. The data from the KaoS
collaboration
are measurements at mid-rapidity extrapolated to total yields under
the assumption of isotropic emission. 
As the beam energy is reduced, there is a steady decrease in the observed
K$^-$/K$^+$ ratio. 

\begin{figure}[htb]
\begin{minipage}[t]{175mm}
\epsfxsize=17.5cm\epsfbox[10 0 670 275]{
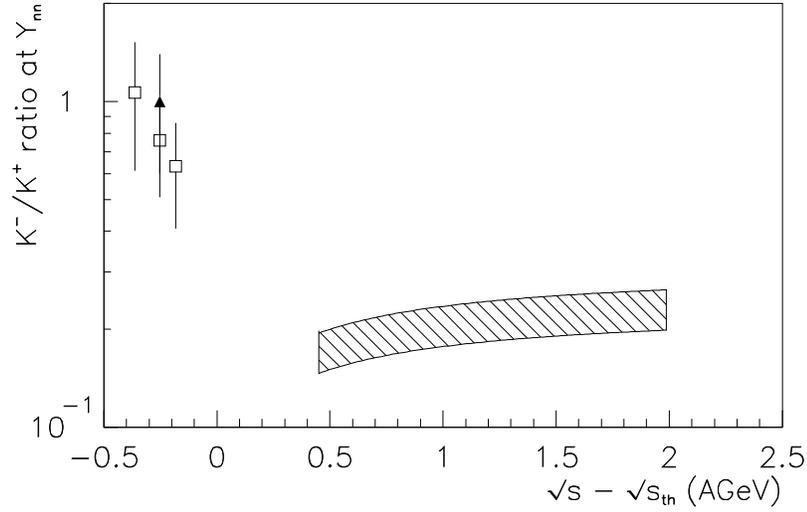}
\end{minipage}

\caption{
The K$^-$/K$^+$ ratio as a function of 
$\protect\sqrt{s}$-$\protect\sqrt{s_{th}}$
The hashed bands represent the data in this letter as $\pm 1\sigma$ bands
calculated from the fitted K$^+$ and
K$^-$ yields for central Au+Au reactions.  
Also shown are the K$^-$/K$^+$ ratios measured by the KaoS collaboration in 
Ni+Ni (triangles)\protect\cite{kaosNiNi} and 
C+C (open squares)\protect\cite{kaosCC}. }
\label{fig:ratiothresh}
\end{figure}

An approach adopted by the KaoS collaboration
(following the work of Shor et al.\cite{Shor} and Metag\cite{Metag}) was
to form the ratio of K$^-$/K$^+$ yields at the same value of
$\sqrt{s}$ - $\sqrt{s_{th}}$ for each species.
The corresponding beam energies are called ``equivalent energies''.
This provides a comparison of the kaon excitation function shapes
with respect to the p+p energy threshold.
The K$^-$/K$^+$ ratio at the same
$\sqrt{s}$ - $\sqrt{s_{th}}$
was directly measured by the KaoS collaboration\cite{kaosNiNi,kaosCC}
at energies below the threshold. 
For the above threshold
data in this Letter, the 
K$^-$/K$^+$ ratio as a function of $\sqrt{s}$ - $\sqrt{s_{th}}$ 
shown in Figure~\ref{fig:ratiothresh}  
was obtained by dividing the 
functions used to fit the K$^-$ and K$^+$ yields
(Figure~\ref{fig:yieldexcit}).  The systematic uncertainty of
the ratio of interpolated kaon yields is estimated to be 10\%.

There is a notable difference between
the K$^-$/K$^+$ ratio at equivalent energy
below threshold where the ratio is approximately
unity\cite{kaosNiNi,kaosCC}, to 
energies above threshold, where we observe the factor of five smaller
ratio K$^-$/K$^+$=0.2.
This demonstrates that the shapes of the excitation functions with
respect to the p+p threshold are different for K$^-$ and K$^+$, with 
the K$^-$ excitation function falling less steeply with decreasing energy.

There are some caveats to this observation.
The collisions studied at lower energies use smaller heavy-ions 
(C and Ni) than the collisions at the higher energies (Au). 
However, for both Ni+Ni collisions at 
low energy\cite{kaosNiNi} and Au+Au collisions at
high-energy\cite{kpkm} 
the K$^-$/K$^+$ ratio has been measured to be independent of 
the centrality of the collision. 
It is also noted that
is no well-defined threshold for kaon production in heavy-ion
collisions, and that in heavy-ion collisions at these energies 
the production of kaons is 
dominated by multiple-scattering of hadrons. 
Therefore shifting the energy scale by $\sqrt{s_{th}}$ may be 
inappropriate for heavy-ion data.
     
There are at least two plausible explanations for the observed difference in 
the shape between
the K$^-$ and K$^+$ excitation functions:
\begin{enumerate}  
\item  At beam energies below 2~AGeV, the measured
ratio of K$^-$/K$^+$ at a fixed $\sqrt{s}$ is less than
0.03 (Figure~\ref{fig:ratioexcit}). If only a 
few percent of the  hyperons (e.g. $\Lambda$ or $\Sigma$) formed in association
with K$^+$ undergo a strangeness transfer
reaction hyperon+hadron$\rightarrow N+K^-$+X , 
this reaction could be the dominant source of 
K$^-$\cite{JSB99,Ko83,Barz85}.
This mechanism is likely to fall less steeply with beam energy than 
K$^-$K$^+$ pair production.
\item  The increased K$^-$/K$^+$ ratio might indicate that 
kaons have modified 
in-medium properties (mass and/or production cross-section)
and that these changes have the greatest 
influence on the production rates near the K$^-$ production
threshold.
\end{enumerate}
These two possibilities illustrate the difficulty of inferring
in-medium properties solely from 
measured yields, as it is first necessary to
fully understand the detailed effects of hadronic rescattering on kaon
production. The
data set in this Letter, coupled with the published pion, and proton
spectra and yields at AGS energies\cite{kpkm,Back99,Chen98} can be used 
to constrain the modelling 
of rescattering in transport calculations, thereby
creating a firmer basis from which to explore the 
physics of in-medium masses. 

A useful step has been made in a recent theoretical paper by Cassing 
et al.\cite{Cass00} which contains predicted total yields of particles 
produced in Au+Au collisions from 1-200 AGeV. Comparisons between data and
calculated yields at mid-rapidity will be made in a later publication.

This Letter has reported on the mid-rapidity spectra and yields of 
K$^-$ from Au+Au
collisions at 4, 6, 8 and 10.7~AGeV. The K$^-$ data have been
compared with K$^+$ data taken at the same energies (and at 2 AGeV)
with the same apparatus.
The K$^+$ and K$^-$ yields increase with beam-energy but not at the same
rate, such that the ratio
K$^-$/K$^+$ increases with increasing beam energy. 
This continues the trend observed 
with the lower energy C+C and Ni+Ni  K$^-$/K$^+$ data taken at GSI. 
To compare 
the yields with respect to the production threshold in p+p reactions,
the ratio K$^-$/K$^+$ is formed with both species evaluated
at the same equivalent energy, 
$\sqrt{s}$-$\sqrt{s_{th}}$.
Data from GSI indicate that the K$^-$/K$^+$ ratio is near unity at
equivalent energies below threshold, 
whereas for the data in this Letter above threshold, the ratio is smaller by
a factor of five, i.e. K$^-$/K$^+$$=0.2$.
This change implies that the yield of K$^-$
in heavy-ion collisions falls less steeply near the nucleon-nucleon
threshold than the yield of
K$^+$.

This work was supported by the U.S. Department of Energy under contracts
with 
ANL (W-31-109-ENG-38),
BNL (DE-AC02-98CH10886), and grants to the
University of California, Riverside (DE-FG03-86ER40271),
Columbia University (DE-FG02-86-ER40281),
University of Illinois, Chicago (DE-FG02-94ER40865),
LLNL (W-7405-ENG-48), and
MIT (DE-FC02-94ER40818),
the U.S. National
Science Foundation under a grant with
University of Rochester (PHY-9722606),
U.S. NASA under contract with the
University of California (NGR-05-003-513),
Ministry of Education, Science, Sports and 
Culture (Japan), and by Korea Research Foundation (1997-001-D00118).

\clearpage
\begin{table}[hbt]
\squeezetable
\begin{tabular}{|c|c|c|c|c|c|c|c|c|c|} 
\hline
E$_{kin}$& y$_{nn}$& K$^+$ dN/dy& K$^+$ $T$& K$^+$$<m_t>$-m$_0$& 
$\chi^2$/d.o.f.
& K$^-$ dN/dy & K$^-$ $T$  &K$^-$ $<m_t>$-m$_0$ & $\chi^2$/d.o.f.\\ 
(A~GeV) & & & (GeV/c$^2$) &  (GeV/c$^2$) & & 
& (GeV/c$^2$) &  (GeV/c$^2$) & \\ \hline
2 & 0.90 & 0.381 $\pm$ 0.015 & 0.138 $\pm$ 0.004 & 0.168 $\pm$ 0.005 &20/26
&                 &                   &                   &     \\
4 & 1.17 & 2.34 $\pm$ 0.05 & 0.158 $\pm$ 0.003 &  0.197 $\pm$ 0.005 &13/18
& 0.19 $\pm$ 0.01 & 0.149 $\pm$ 0.008 & 0.183 $\pm$ 0.011 &17/19 \\
6 & 1.34 & 4.84 $\pm$ 0.09 & 0.208 $\pm$ 0.006  & 0.270 $\pm$ 0.009 &9/14
& 0.61 $\pm$ 0.02 & 0.166 $\pm$ 0.010 & 0.208 $\pm$ 0.014 &5/14 \\
8 & 1.47 & 7.85 $\pm$ 0.21 & 0.220 $\pm$ 0.011 & 0.287 $\pm$ 0.017 &7/9
& 1.26 $\pm$ 0.04 & 0.173 $\pm$ 0.009 & 0.218 $\pm$ 0.013 &4/10 \\
10.7 & 1.60 & 11.55 $\pm$ 0.24 & 0.204 $\pm$ 0.006 & 0.264 $\pm$ 0.008 &15/18
& 2.21 $\pm$ 0.03 & 0.200 $\pm$ 0.003 & 0.258 $\pm$ 0.005 &19/18 \\
\hline 
\end{tabular}
\parbox{15.0cm}{\caption [] {\label{tabk0p5}
Excitation function of spectral characteristics for   
mid-rapidity K$^+$ and K$^-$ from the most central 5\% Au+Au reactions.                
The mid-rapidity range 
for 2, 4, 6, 8 AGeV is $|\frac{y-y_{nn}}{y_{nn}}|<0.25$,
for 10.7 AGeV the width is  $|\frac{y-y_{nn}}{y_{nn}}|<0.125$,
where $y_{nn}$ is  mid-rapidity in the laboratory frame.  
The errors are statistical only.
}}
\end{table}
\clearpage

\clearpage           

\clearpage

\clearpage
\end{document}